\documentclass[%
groupedaddress,
nofootinbib,
 amsmath,amssymb,
 aps, %physrev,
 prl,
 reprint,
]{revtex4-2}

\usepackage{color}             %for \textcolor
\usepackage{csquotes}        %enquote
\usepackage{pgfplots,float}
\usepackage{caption}
\usepackage{subcaption}
\usepackage{ulem}
\definecolor{tab10Green}{rgb}{0.1726,0.6275,0.1726}
\definecolor{tab10Red}{rgb}{0.8392,0.1530,0.1569}
\definecolor{tab10Blue}{rgb}{0.1216,0.4667,0.7059}
\definecolor{tab10Orange}{RGB}{255,127,14}
\definecolor{tab10Gray}{RGB}{127,127,127}
\usepackage{graphicx}% Include figure files
\usepackage{dcolumn}% Align table columns on decimal point
\usepackage{bm}% bold math
\usepackage{comment}

\usepackage[%
    pdfusetitle,
    colorlinks=true,
    citecolor=tab10Green,
    linkcolor=tab10Red,
    urlcolor=tab10Blue,
]{hyperref}% add hypertext capabilities
\usepackage{braket}            %for braket notation
\newcommand{\nt}{N_\tau}
\newcommand{\meff}{\ensuremath{M_{\text{eff}}}}
\DeclareMathOperator{\Tr}{Tr}

\begin{document}

\author{Rachel Horohan D'Arcy}
\affiliation{
 Department of Physics, National University of Ireland Maynooth, County Kildare, Ireland
}
\author{Benjamin J\"ager}
\affiliation{%
 Quantum Field Theory Center \& Danish IAS, Department of Mathematics and Computer Science \\
  University of Southern Denmark, 5230, Odense M, Denmark
}%
\author{Seyong Kim}
\affiliation{
Department of Physics, Sejong University, Seoul 05006, Korea
}%
\author{Jon-Ivar Skullerud}
\affiliation{
 Department of Physics, National University of Ireland Maynooth, County Kildare, Ireland
}

\title{Thermal modifications of the B-meson spectrum}
\date{\today}
\begin{abstract}
A first-principles investigation of heavy-light meson behaviour in hot QCD matter is essential for the interpretation of experimental results associated with open heavy flavour productions in relativistic heavy-ion collision experiments.  Using anisotropic lattice ensembles from the \textsc{Fastsum} collaboration with $N_f = 2 + 1$ dynamical Wilson-clover fermions at non-zero temperatures, we study the $B$ and $B_s$ spectra at non-zero temperature.  We use relativistic light quark propagators, while the $b$ quark propagators are computed with a non-relativistic effective theory (NRQCD).  We find that above $T_c$, thermal effects on the $B$ meson states are more significant than those on the $B_s$ states. Also, our results support the dissolution of the $B$-meson bound states above $T_c$.   
\end{abstract}
\maketitle

\paragraph{Introduction} -
Understanding the behaviour of QCD matter in extreme conditions is one of key objectives in modern nuclear and particle physics. At high temperatures strongly interacting matter is expected to undergo a transition from confined hadrons to a medium of deconfined quarks and gluons, the quark-gluon plasma (QGP).  This medium is studied in heavy-ion collisions at the Relativistic Heavy Ion Collider (RHIC) and the Large Hadron Collider (LHC). Hard probes such as high-$p_T$ jets and heavy quarks play a key role in the exploration of the QGP because they are produced in early stages of collisions and then propagate through the medium \cite{Apolinario:2022vzg}.

In this work, we focus on heavy-flavour mesons which are particularly useful probes into the non-perturbative regime of QCD \cite{Andronic:2015wma,Dong:2019byy}. Both open-heavy flavour mesons and quarkonia involve heavy quarks but explore different aspects of the QGP phase. In particular, open-heavy flavour mesons provide insights on the heavy quark diffusion and about the hadronisation process at freeze-out, while the relative yields of quarkonia and open-heavy flavour are invaluable to understanding the fate of heavy quarks in the medium.  

There has been considerable theoretical effort devoted to the study of quarkonia at high temperature \cite{Rapp:2008tf,Rothkopf:2019ipj,Andronic:2024oxz}, and more recently to open-charm mesons \cite{Kelly:2018hsi,Aarts:2022krz,Das:2024vac}. In contrast, there is comparatively little known about open-beauty quark mesons, both theoretically \cite{Torres-Rincon:2026rso} and experimentally \cite{CMS:2017uoy,CMS:2021mzx,CMS:2024vip}. In particular, whether or not B-mesons survive above the chiral crossover temperature remains an open question.  
Lattice QCD offers a non-perturbative framework to study strongly interacting matter in thermal equilibrium. Euclidean correlation functions are calculated on the lattice from which hadronic properties can be extracted. This extraction is non-trivial due to the need to indirectly infer the spectral information from Euclidean time correlators. Despite this, lattice QCD is the only systematically improvable approach to investigate the behaviour of hadronic states at high temperature. 

In this study we discuss the effects of temperature on the spectrum of $B$ and $B_s$ mesons in the pseudoscalar and vector channels, using lattice QCD with relativistic light and strange quarks and non-relativistic QCD (NRQCD) for the beauty quarks. We use standard effective mass fits to determine the mass spectrum at finite temperature. We then use spectral reconstruction to further understand the in-medium behaviour. 
We find significant thermal effects above the chiral crossover temperature $T_c$, with more significant effects felt by the $B$ compared to the $B_s$. Our results suggest that $B$-meson bound states do not survive above $T_c$.

\paragraph{Formalism} -
 Direct simulation of $b$ quarks requires extremely fine lattices to satisfy $a_\tau m_b \ll 1$, making simulations computationally expensive. To avoid this, we use NRQCD \cite{Thacker:1990bm,Lepage:1992tx,Davies:1994mp}, an effective field theory used to describe heavy quarks in QCD. 
The relativistic modes are removed by integrating out the UV degrees of freedom with $p > 2M$, leaving the non-relativistic effective theory. The action is then expanded in powers of the heavy quark velocity ($v \ll 1$) and the Dirac field is transformed to decouple the upper and lower components, separating the quark and anti-quark fields. The time evolution becomes an initial value problem with only forward in time propagation of the quarks, a significant simplification over the relativistic approach.  It is worth noting that the heavy quarks move even more slowly in open-heavy-flavour systems than in quarkonia: $v^2/c^2 \approx 0.1$ in $b\bar{b}$ systems and $v^2/c^2 \approx 0.01$ in B-mesons, where $v$ is the heavy quark velocity in the rest frame of the respective meson.

We constructed two-point correlation functions for B-mesons by combining NRQCD propagators for the beauty anti-quark with relativistic propagators for the light quark. The light quark propagator is computed
with anti-periodic boundary conditions in time, while the NRQCD heavy quark field is decoupled and satisfies an initial condition problem, and therefore only the backward anti-quark propagator is calculated.

The two-point correlator $G(\tau)$ is given by
\begin{align}
G(\tau) &= \sum_{\vec{x}}\braket{\bar\Psi_q(\vec{x},\tau)\Gamma\Psi_Q(\vec{x},\tau)\bar\Psi_Q(0,0)\Gamma\Psi_q(0,0)} \\
    & = - \sum_{\vec{x}}\Tr\big[ \Gamma S_q(\vec{x},\tau;0,0)
    \Gamma S_Q(0,0;\vec{x},\tau)\big], \label{eq:TrGSGS}
\end{align}
where the subscripts $q$ and $Q$ refer to the light and heavy quark respectively, and the trace runs over the colour indices and the (upper) two spin indices \cite{Gregory:2010gm}. 
The $\Gamma$ matrices in Eq.~\eqref{eq:TrGSGS} are $I_{2\times2}$ and $\vec{\sigma}$ for the pseudoscalar and vector mesons respectively.

\paragraph{Lattice Details ---}
We use \textsc{Fastsum}'s Generation 2 and 2L (Gen2/2L) anisotropic lattice ensembles, with a Symanzik-improved anisotropic gauge action and $N_f=2+1$ dynamical quark flavours with an $\mathcal{O}(a)$-improved Wilson fermion action. The strange quark mass is tuned to be approximately physical while the light quarks are heavier than physical, with $M_\pi = 390$ MeV and $M_\pi = 240$ MeV for Gen2 and Gen2L, respectively, see Table~\ref{tab:latticeparams}. Anisotropic lattices provide a fine temporal resolution, essential for a reliable determination of spectral properties, with moderate computational cost. We employ a fixed scale approach where the temperature $T=(a_\tau N_\tau)^{-1}$ is varied by changing the number of temporal lattice sites $N_\tau$, see Table \ref{tab:Temperatures}. For details of the ensembles see Refs \cite{Aarts:2014nba, Aarts:2020vyb, Aarts:2022krz}. 

The use of NRQCD entails the use of an additive mass shift related to the rest mass of the $b$ quark. This energy shift is determined using the spin-averaged $b\bar{b}$ 1S mass.  For for the B-mesons we take the mass shift to be half that of the $b\bar{b}$ sector, giving $\Delta E = 4126$ MeV (Gen2) \cite{Aarts:2014cda} and $\Delta E = 3732$ MeV (Gen2L) \cite{NRQCD-Gen2L}.
\begin{table}[t]
    \centering
    \begin{tabular}{c|cccccc}
          Gen  & $\xi$ & $a_s$ (fm) & $a_\tau^{-1}$(Gev) & $m_\pi$(MeV) & $N_s$ & $T_c$ \\ \hline
            2  & 3.444(6) & 0.1205(8) & 5.63(4) & 384(4) & 24 & 181(1) \\
            2L & 3.453(6) & 0.1121(3) & 6.079(13) & 239(1) & 32 & 167(3) 
        \end{tabular}
        \caption{Lattice parameters for \textsc{Fastsum} ensembles Gen2 and Gen2L: Spatial lattice spacing $a_s$, inverse temporal lattice spacing $a_\tau^{-1}$, anisotropy $\xi=a_s / a_\tau$, pion mass $m_\pi$, number of lattice sites in the spatial direction $N_s$, and the pseudocritical temperature $T_c$.}
    \label{tab:latticeparams} 
\end{table}
\begin{table}[t]
    \centering
    \begin{tabular*}{0.7\linewidth}{@{\extracolsep{\fill}} c|cc|cc}
                & Gen2 & & Gen2L &  \\ \hline
        $N_\tau$& $T$ (MeV) & $T/T_c$ & $T$ (MeV) & $T/T_c$ \\ \hline
            128 & 44  & 0.24 & 47  & 0.28 \\
            64  & --  &  --  & 95  & 0.57 \\
            56  & --  &  --  & 109 & 0.65 \\
            48  & 117 & 0.65 & 127 & 0.76 \\
            40  & 141 & 0.78 & 152 & 0.91 \\
            36  & 156 & 0.86 & 169 & 1.01 \\
            32  & 176 & 0.97 & 190 & 1.14 \\
            28  & 201 & 1.11 & 217 & 1.30 \\
            24  & 235 & 1.30 & 253 & 1.52 \\
    \end{tabular*}
    \caption{Temporal lattice extents $\nt$ and corresponding temperatures $T$ used in this study, in units of MeV and the chiral pseudocritical temperature $T_c$.}
    \label{tab:Temperatures}
\end{table}

\paragraph{Zero-temperature results ---}
\begin{table}[t]
    \centering
    \begin{tabular}{c|cc|cc|c}
                   % & Gen2 & &Gen2L  &&PDG \cite{ParticleDataGroup:2026aaa} \\ \hline
                    & $M_{\text{eff}}$ & &  BR& & \\ \hline
                    & Gen2 & Gen2L &Gen2  & Gen2L&PDG \cite{ParticleDataGroup:2026aaa} \\ \hline
    $B$           & 5335(41) & 5338(16) & 5319(40) & 5318(18) & 5279.72(8)  \\
    $B^*$         & 5376(42) & 5368(17) & 5392(43) & 5374(20) & 5324.75(20) \\
    $B_s$         & 5386(41) & 5420(14) & 5374(40) & 5406(24) & 5366.93(10) \\
    $B^*_s$       & 5435(41) & 5454(14) & 5428(41) & 5440(13) & 5415.4(1.4) \\ \hline
    $B^* - B$   & 41(20)   & 30(20)   & 72(44)   & 55(18)   & 45.18(20)   \\
    $B_s - B$     & 51(18)   & 82(18)   & 55(41)   & 87(18)   & 87.45(44)   \\ 
    $B^*_s - B_s$ & 47(17)   & 33(15)   & 54(40)   & 35(14)   & 48.5(1.4)   \\ 
    \end{tabular}
    \caption{Ground state masses and mass splittings in MeV at zero temperature, from effective mass fits (\meff) and BR spectral function reconstruction (BR), compared to experimental values (PDG).}
    \label{tab:T0fit&BRvalues}
\end{table}
Before we discuss the thermal modifications we will present the zero-temperature, $T_0$, results, obtained from the $\nt=128$ lattices. Table \ref{tab:T0fit&BRvalues} shows the masses and mass splittings obtained from effective mass fits (detailed below) at $T=0$ compared to the experimental values. The difference between the our results and the PDG \cite{ParticleDataGroup:2026aaa} can be partly explained by the tuning of the light quarks to heavier than physical, while the strange quark mass is tuned to be approximately physical.   Our results for Gen2 and Gen2L are in agreement for the for the $B$ and $B_s$, and the various mass splittings. The mass splittings for Gen2 and Gen2L agree with each other and are also in excellent agreement with the experimental values. We take this as confirmation that the systematics of our hybrid relativistic + NRQCD approach is under control.  We note that our main interest here is in how the states change with temperature and not in a precision study of the zero-temperature spectrum.

As is detailed below, we also use the Bayesian Reconstruction (BR) method, to find the spectral function, $\rho(\omega)$, of the B-mesons at each temperature. We identify the peak position as the ground state energy, and these results are also shown in Table \ref{tab:T0fit&BRvalues}. We see that both methods give consistent results, giving further confidence in our approach.

\paragraph{Correlators ---}
The B-meson correlator \eqref{eq:TrGSGS} is constructed from a light quark propagator containing both forward and backward propagation, and the NRQCD heavy (anti-)quark propagator, which is only forward propagating. To avoid any contamination from the backward propagating light quark, we will in the following use only temporal separations $\tau/a_\tau<N_\tau/2$. We have checked that our results are robust against small variations in this cutoff.

\begin{figure}[t]
    \centering
   \includegraphics[width=0.9\linewidth]{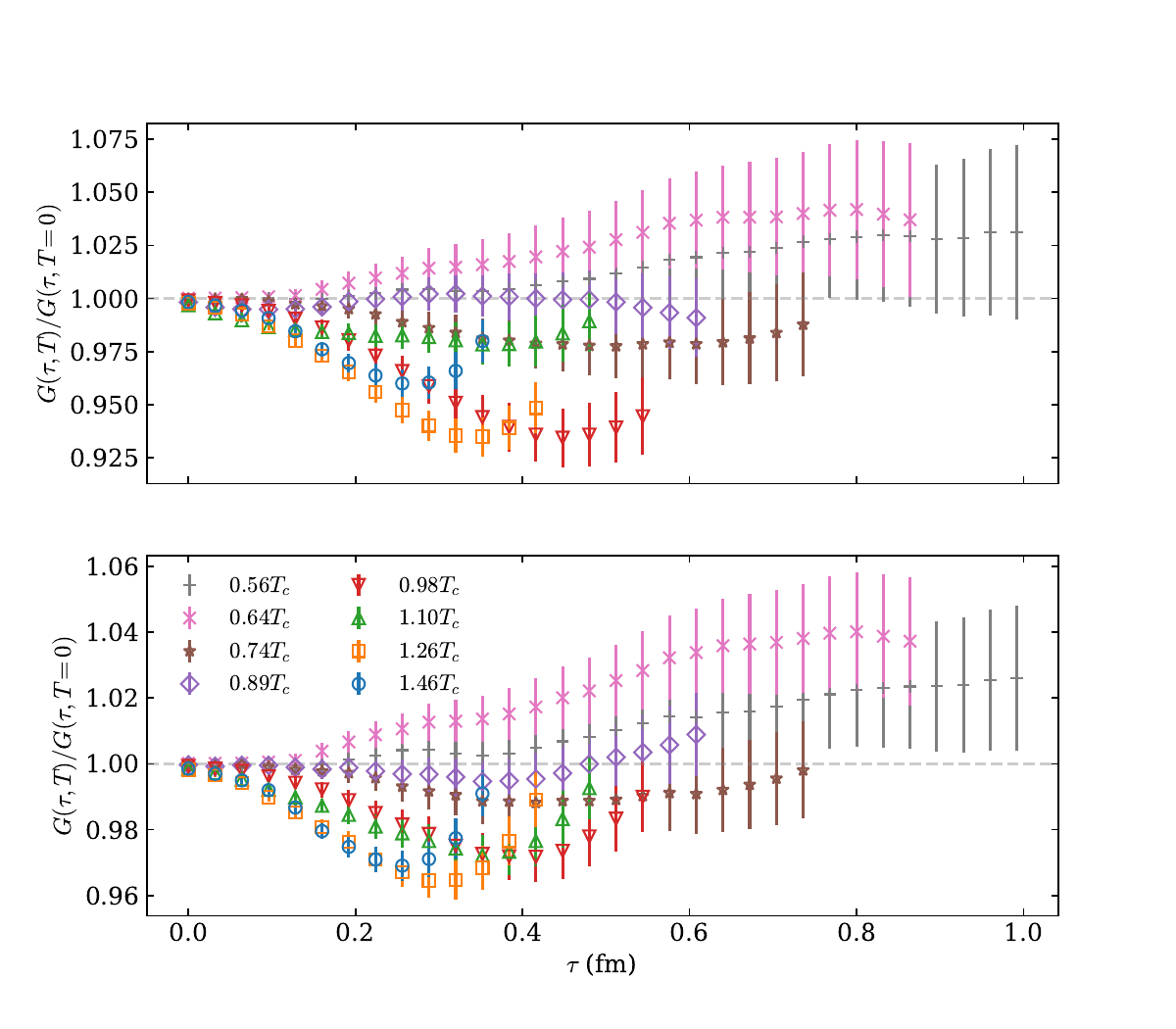}
    \caption{Ratio of the in medium correlators to the zero-temperature correlator, for $B$ (top) and $B_s$ (bottom) from Gen2L.% (Note we only show half the temporal extent.)
    }
    \label{fig:corrratios}
\end{figure}

For a first indication of thermal effects, we show the ratio of the thermal correlators and the zero-temperature correlator in Fig. \ref{fig:corrratios}. This ratio is equal to 1 in the absence of thermal effects, meaning that any significant deviation from 1 will be a sign of  thermal modifications. We see that there is no statistically significant deviation from 1 for $T < T_c$, indicating that there are few to no thermal effects at these temperatures. For $T > T_c$ there is a significant deviation from 1, giving us the first indication of in-medium modifications to the correlator. There is a significant effect at $T \approx T_c$ for both the $B$ and $B_s$ but the $B$ has a more prominent deviation from 1. We only show the ratio for the pseudoscalar from Gen2L but found the same qualitative effects in the pseudoscalar and vector channels from both ensembles. We find that the $B$ exhibits more significant thermal effects than the $B_s$ in both Gen2 and Gen2L lattice ensembles. 

\paragraph{Effective mass fits ---}
To extract the temperature-dependent mass of the B-mesons we fit the standard effective mass, 
\begin{gather}
   a_\tau M_{\text{eff}}(\tau) = \log\left(\frac{G(\tau)}{G(\tau + a_\tau)}\right), \label{eq:Meff}\\
    \intertext{to the form,}
    M_{\text{eff}}(\tau) = M + Ae^{-b\tau},\label{eq:effmassfit}
\end{gather}
where $M$ is the ground state mass while the second term accounts for the exponential suppression of the excited states.  
In the fixed-scale approach, the number of temporal points is inversely proportional to the temperature, and we only use half those available points in this study. To distinguish between the effect that a limited number of points can have on the result and actual thermal effects we do a zero-temperature ($T_0$) analysis. We truncate the zero-temperature ($N_\tau = 128$) correlators to equal the temporal extent of the higher temperatures, and repeat the analysis on these truncated correlators. The true thermal effect is then taken to be the difference between the two results. 

\begin{figure}[t]
    \centering
    \includegraphics[width=0.8\linewidth]{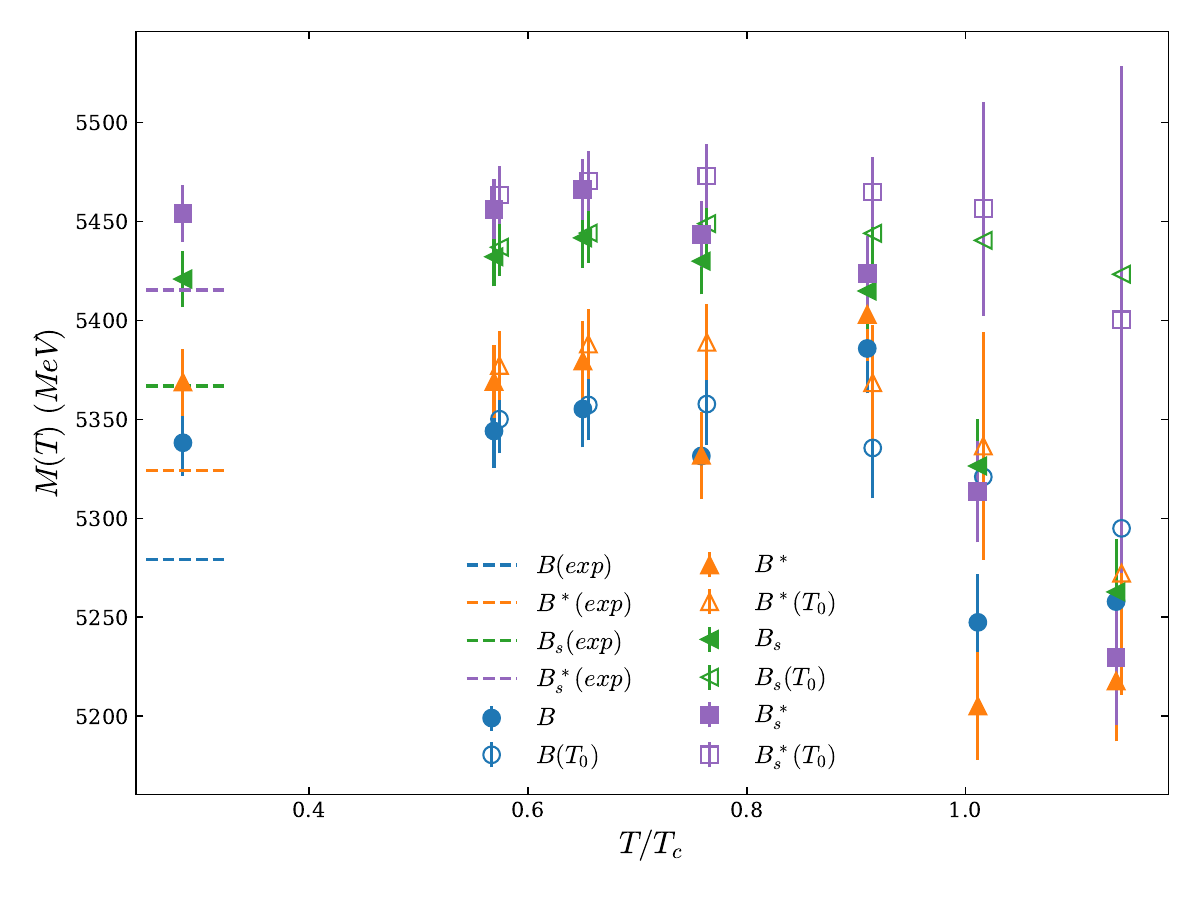}
    \caption{Effective mass fits for the in medium (filled markers) $B$, $B^*$, $B_s$, and $B^*_s$ and $T_0$ analysis (empty markers), with the corresponding experimental values \cite{ParticleDataGroup:2026aaa} (dashed lines) for Gen2L.}
    \label{fig:G2lEffmassFits}
\end{figure}

\begin{figure}[t]
    \centering
    \includegraphics[width=0.8\linewidth]{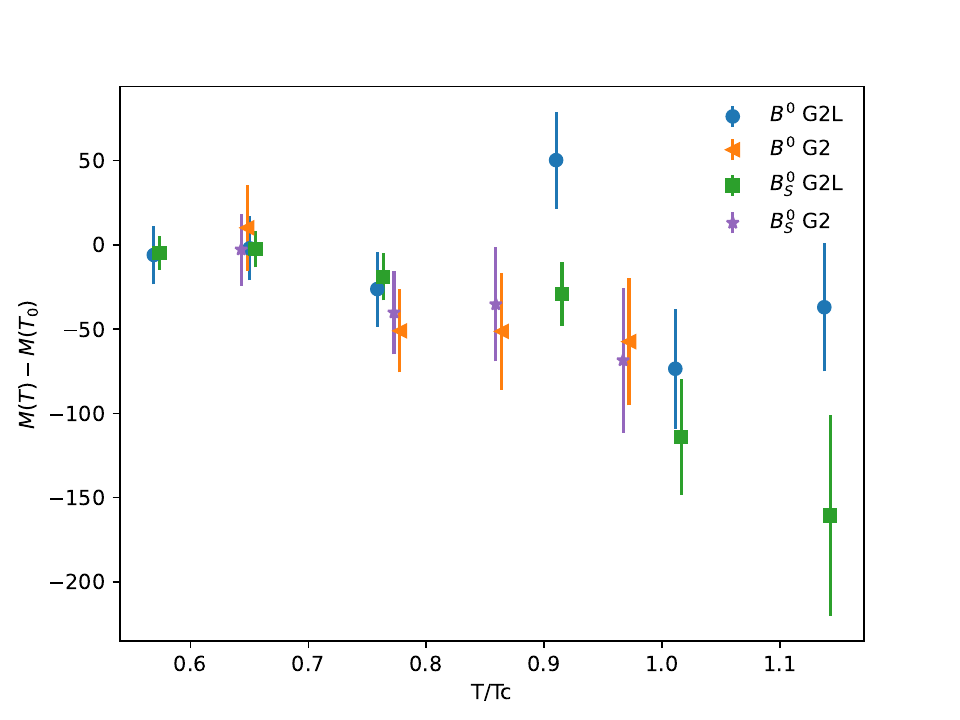}
    \caption{Thermal mass shift for the $B$ and $B_s$ mesons from Gen2 and Gen2L, from effective mass fits.  The mass shift is obtained as the difference between the mass extracted from thermal correlators and the corresponding truncated $T_0$ correlator. We have shifted the points horizontally by small amounts for clarity.}
    \label{fig:deltaMG2G2l_B0}
\end{figure}

The in-medium effective mass fits are qualitatively the same for Gen2 and Gen2L. Figure \ref{fig:G2lEffmassFits} shows the Gen2L results for the in-medium and $T_0$ analysis for all four channels, and Figure \ref{fig:deltaMG2G2l_B0} shows the difference between the thermal and truncated $T_0$ results (for the pseudoscalar only). We begin to see a statistically significant difference for $T\gtrsim0.7T_c$, where we find a small negative mass shift that increases with increasing temperature. These results are in quantitative agreement with a recent study done using a thermal EFT approach based on chiral perturbation theory \cite{Montana:2023sft}. 

The error bars shown are a combination of statistical uncertainties estimated using bootstrap sampling, and systematic uncertainties determined using AIC  \cite{Akaike:1998zah} model averaging to average over different fit windows.

\paragraph{Spectral functions ---}
The spectral function $\rho(\omega,T)$ is related to the Euclidean correlator $G(\tau,T)$ according to
\begin{equation}
    G(\tau,T) = \int K(\omega,\tau) \rho(\omega,T)d\omega , 
\end{equation}
where we take $K(\omega,\tau) = e^{-\omega\tau}$, which is justified by us only considering $\tau<1/2T$.
Many different methods have been devised to attempt to solve the ill-posed problem of reconstructing $\rho$ given discrete, noisy data from $G$, and a comparison of some of these methods can be found in Refs. \cite{Spriggs:2021dsb,Skullerud:2025iqt}. Here we use Bayesian Reconstruction (BR) \cite{Burnier:2013nla} to estimate the spectral function, a method that employs Bayesian inference and a default model encoding prior information to perform the reconstruction. 
We repeat the truncated zero-temperature analysis as it has been shown that varying the $\tau$ range can affect the resulting spectral function \cite{Kelly:2018hsi,Rothkopf:2022ctl}. 

\begin{figure}[t]
    \begin{subfigure}[b]{0.95\linewidth}
    \centering
    \includegraphics[width=0.9\linewidth]{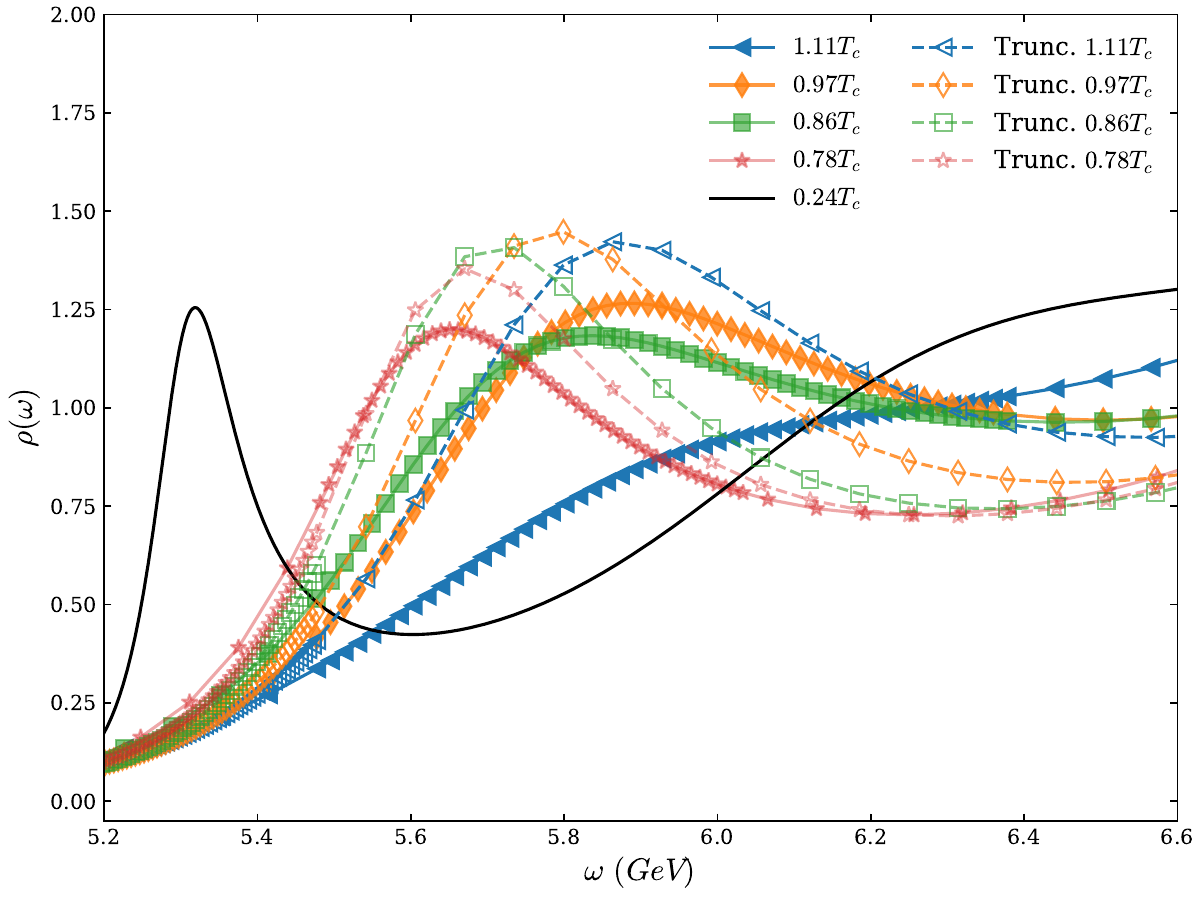}
    \end{subfigure}
    \begin{subfigure}[b]{0.95\linewidth}
    \centering
    \includegraphics[width=0.9\linewidth]{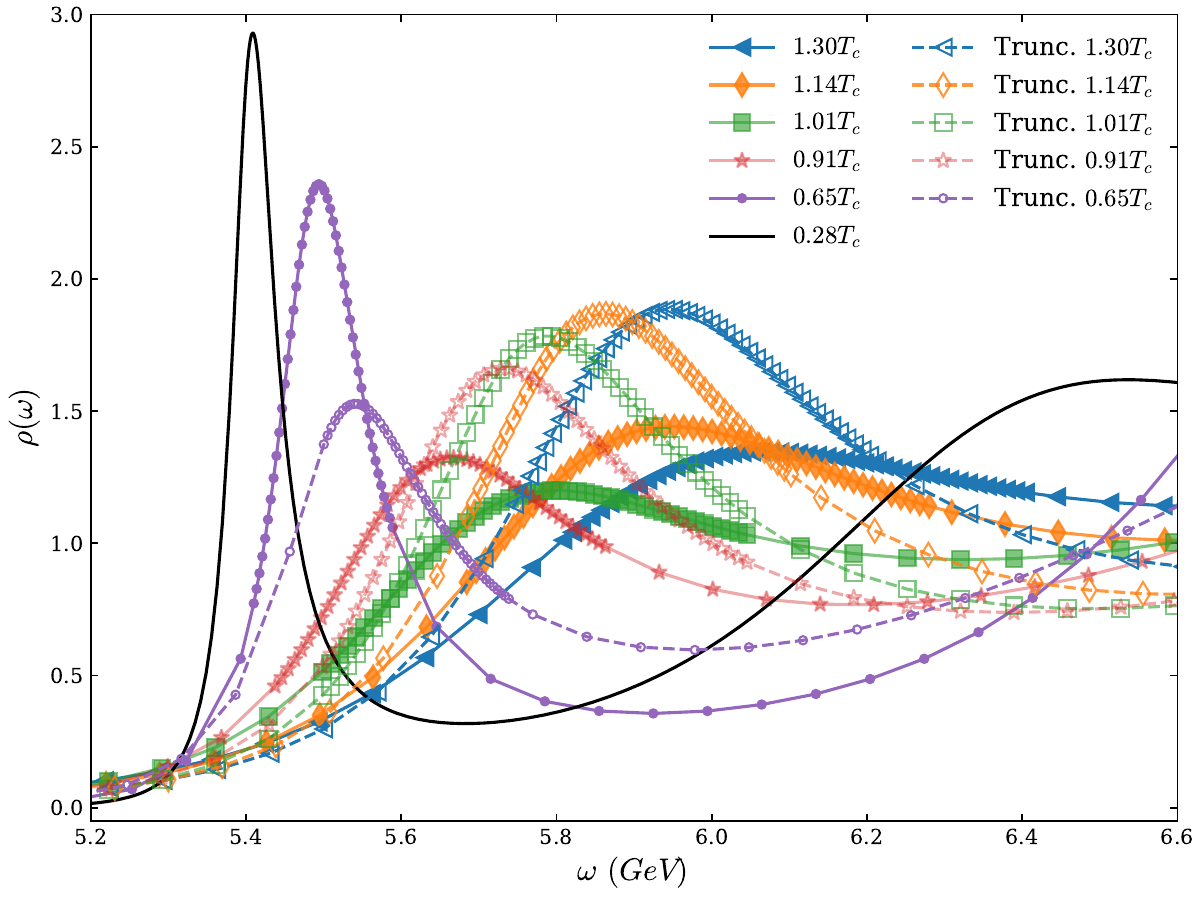}
    \end{subfigure}
    \caption{Spectral functions for the in medium (filled markers) Gen2 $B$ (top) and Gen2L $B_s$ (bottom), and corresponding $T_0$ data (empty markers) from BR. Shown is every 10th point of the reconstructed spectral function.}
    \label{fig:specfuncs}
\end{figure}
We find that the ground state peak disappears above $T_c$ for the $B$ and $B^*$, but a small, very broad bump remains for the $B_s$ and $B_s^*$. Figure \ref{fig:specfuncs} shows the Gen2 $B$ and Gen2L $B_s$ spectral functions for selected temperatures together with the corresponding truncated $T=0$ spectral functions. We can see that the broadening of the peak is not due to the number of temporal points since the $T_0$ spectral function in all cases has a defined peak at $\nt$ equivalent to higher temperatures. This suggests that no bound state remains above $T_c$. 

We have determined the thermal mass shift as as the difference between the thermal ground state peak and the truncated $T_0$ ground state peak, with statistical uncertainties estimated using a jackknife analysis on the spectral reconstruction with 10 jackknife bins, and systematic uncertainties estimated by varying the default model and the $\tau$ and $\omega$ ranges. The results 
are consistent with those from the effective mass fits below $T_c$, but with much larger errors (dominated by the systematic uncertainties). 

\paragraph{Conclusions ---}
We have presented the first calculation of
the spectrum of open-beauty mesons at finite temperature using lattice QCD.  Below the chiral crossover temperature $T_c$ we find a small, negative mass shift for $T\gtrsim0.7T_c$, in agreement with the chiral EFT results found in Ref \cite{Montana:2023sft}, and in qualitative agreement with previous studies of open-charm mesons on the same lattice ensembles \cite{Kelly:2018hsi,Aarts:2022krz}.  Spectral functions obtained using Bayesian Reconstruction suggest no bound states survives above $T_c$, although the results are consistent with $B_s$ mesons surviving up to $T\sim1.3T_c$.  Our correlator and spectral function results suggest that the $B_s$ mesons experience milder thermal modifications than the $B$ mesons, which may be relevant for the $B_s/B$ observed experimentally \cite{CMS:2021mzx}.  

This study has been carried out using a single lattice spacing, with heavier-than-physical light quarks.  To investigate the associated systematics, we plan to repeat this study on our new Generation 3 ensembles \cite{Skullerud:2025xva}, which have twice the temporal resolution, and also on lattice ensembles with physical quark masses, which will be available in the near future.  Work is also in progress to compute $B$ meson properties using fully relativistic $b$ quarks \cite{Gayer:2024akw}.

\paragraph{Open access statement ---}
For the purpose of open
access, the authors have applied a Creative Commons
Attribution (CC BY) licence to any Author Accepted
Manuscript version arising.

\paragraph{Software and data ---}

The NRQCD propagators were produced using \textsc{NRQCD-Fastsum} \cite{fastnrqcd}.  Light and strange quark propagators were computed using the \textsc{OpenQCD-propagator} code, which is part of the \textsc{OpenQCD-Fastsum} package \cite{openqcd-fastsum}.  The correlators were calculated with the code in Ref.~\cite{OpenHeavyCorrelators}, where the correlator data and effective mass analysis code can also be found. The BR spectral reconstruction was carried out using the public BR code \cite{BR}.  

\paragraph{Authors' contributions ---}
\begin{itemize}
\item Horohan D'Arcy: Data production, analysis, draft of manuscript
\item J\"ager: Data production
\item Kim: Contribution to manuscript and physics interpretation of results
\item Skullerud: Data production, contribution to manuscript and physics interpretation.
\end{itemize}
\paragraph{Acknowledgments ---} We are grateful to the Hadron Spectrum Collaboration for the use of their zero temperature ensembles. RHD has been supported by Taighde Éireann – Research Ireland under Grant number GOIPG/2024/3507. SK is supported by the National Research Foundation of Korea under grant RS-2008-NR007226 and by the Institute of Information and Communication Technology Planning and Evaluation under grant IITP-2024-RS-2024-00437191, funded by the Korean government (Ministry of Science and ICT). This work is supported by Irish Centre for High End Computing (ICHEC). This work used the DiRAC Data Intensive service (DIaL2 / DIaL [*]) at the University of Leicester, managed by the University of Leicester Research Computing Service on behalf of the STFC DiRAC HPC Facility (www.dirac.ac.uk). The DiRAC service at Leicester was funded by BEIS, UKRI and STFC capital funding and STFC operations grants. DiRAC is part of the UKRI Digital Research Infrastructure.

\clearpage
\section{End Matter}
\paragraph{Spectral function error analysis ---}

\begin{figure*}
    \centering
    \includegraphics[width=0.45\linewidth]{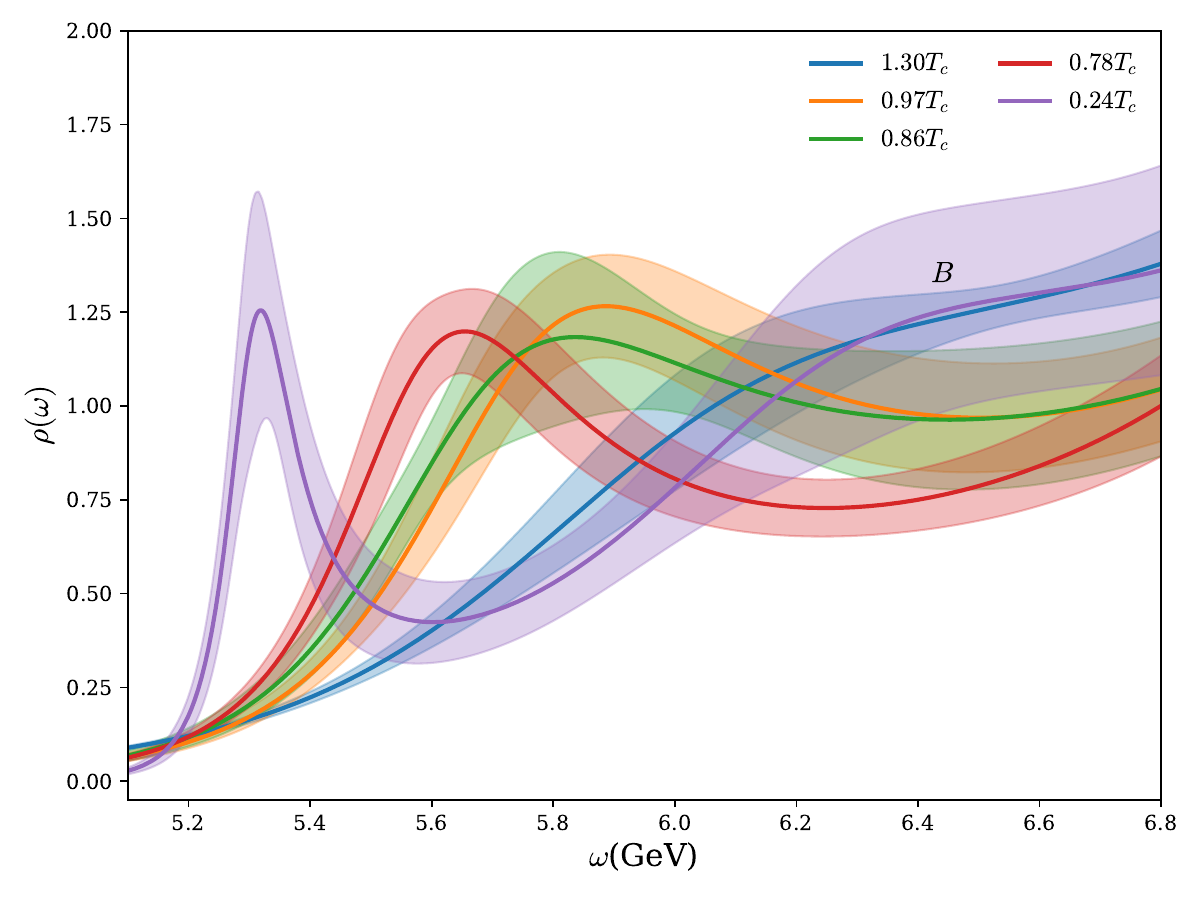}
    \includegraphics[width=0.45\linewidth]{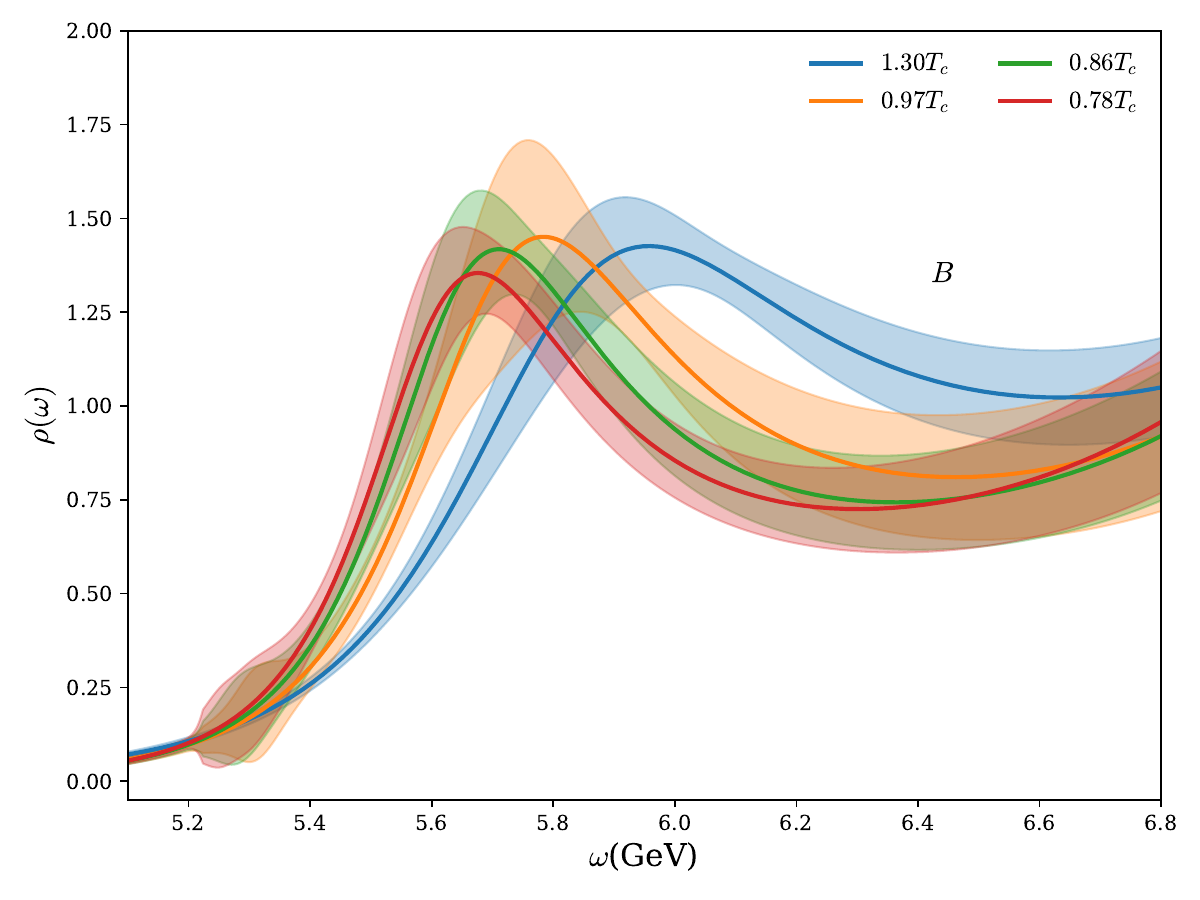}
    \includegraphics[width=0.45\linewidth]{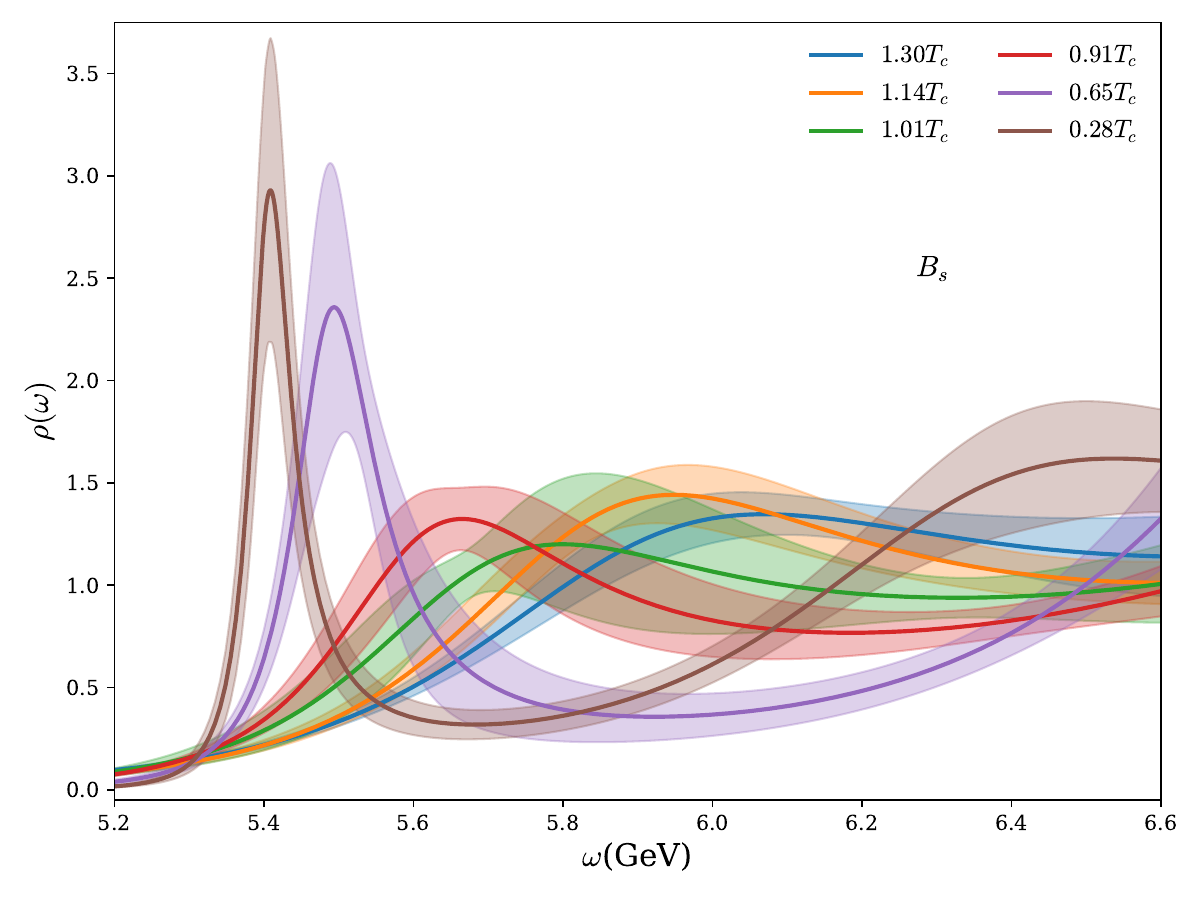}
    \includegraphics[width=0.45\linewidth]{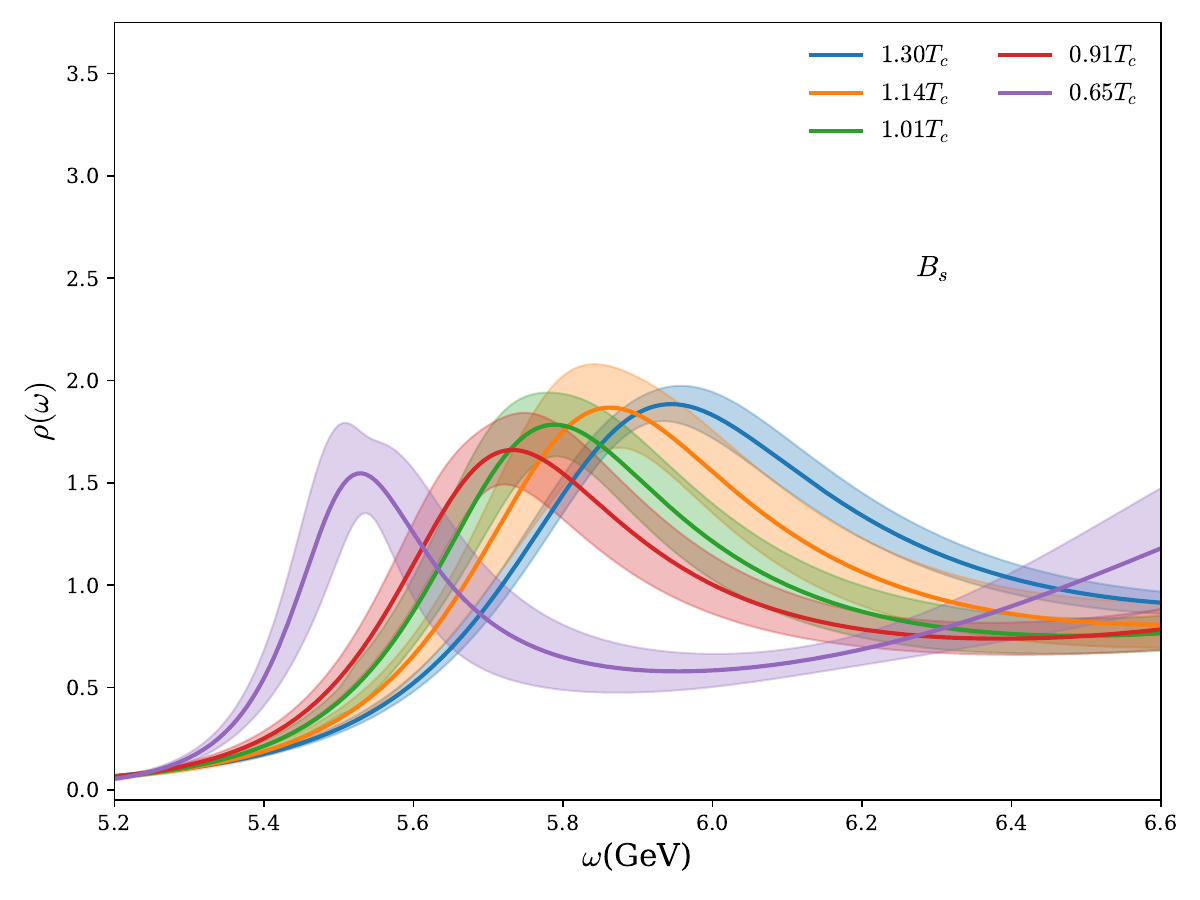}
    \caption{Spectral functions with combined statistical and systematic uncertainties, from thermal (left) and truncated $T_0$ correlators (right).  The top panels show the $B$ channel from Gen2, while the bottom panels show the $B_s$ channel for Gen2L.}
    \label{fig:BR-Bs-errors}
\end{figure*}
The statistical uncertainties in the BR spectral function reconstruction were estimated using a jackknife analysis with 10 jackknife bins. To estimate the systematic uncertainties, we varied the default model, the frequency range, and the imaginary time range.
The central values shown in Fig.~\ref{fig:specfuncs} were obtained using a constant default model, $a_\tau\omega\in[-0.05,4]$ and $\tau/a_\tau\in[2,N_\tau/2-1]$.  The default model $D(\omega)$ was varied using $D(\omega)\propto(\omega-\omega_{\min}+a_\tau^{-1})^n$ with $n=-2,1,2$.  The minimum and maximum frequencies $\omega_{\min},\omega_{\max}$ were varied with $a_\tau\omega_{\min}\in\{-0.5,0,0.05\}$ and $a_\tau\omega_{\max}\in\{4,5,6\}$, while the time ranges were varied with $\tau_{\min}/a_\tau\in\{2,3,4\}$ and $\tau_{\max}/a_\tau\in\{N_\tau/2-1,N_\tau/2-2,N_\tau/2-3\}$.  The spread between the resulting spectral functions (always varying one parameter at a time) was taken as the systematic uncertainty, and combined in quadrature with the statistical uncertainty to yield the error bands shown in Fig.~\ref{fig:BR-Bs-errors} (left) for the $B_s$ channel from Gen2L.

The same error analysis was carried out on the truncated data.  We note that for the data corresponding to $N_\tau \leq 40$ the smallest $\tau_{max}$ is the same as the central $\tau_{max}$ for the next highest temperature.  The resulting spectral functions with error bands for $B_s$ from Gen2L are shown in Fig.~\ref{fig:BR-Bs-errors} (right).  We find that the uncertainties are largely under control and do not affect our main conclusions.
\bibliography{bmesons}
\end{document}